\documentclass[preprint,aps,nofootinbib,epsfig]{revtex4}
\usepackage{amsmath,amssymb,bm}
\usepackage{epsfig}
\usepackage{graphicx}
\usepackage{slashed}
\usepackage{psfig}

\def\lsim{\raise0.3ex\hbox{$<$\kern-0.75em\raise-1.1ex\hbox{$\sim$}}}
\def\gsim{\raise0.3ex\hbox{$>$\kern-0.75em\raise-1.1ex\hbox{$\sim$}}}

\def\pom{{I\!\!P}}
\newcommand{\be}{\begin{equation}}
\newcommand{\ee}{\end{equation}}

\def\beq{\begin{equation}}
\def\eeq{\end{equation}}
\def\beqa{\begin{eqnarray}}
\def\eeqa{\end{eqnarray}}

\newcommand{\rr}{\mbox{\boldmath $r$}}

\newcommand{\rb}{\mbox{\boldmath $b$}}
\newcommand{\calN}{\mathcal{N}}
\def\gappeq{\mathrel{\rlap {\raise.5ex\hbox{$>$}}
{\lower.5ex\hbox{$\sim$}}}}

\def\lappeq{\mathrel{\rlap{\raise.5ex\hbox{$<$}}
{\lower.5ex\hbox{$\sim$}}}}

\def\Toprel#1\over#2{\mathrel{\mathop{#2}\limits^{#1}}}

\newcommand{\rk}{\mbox{\boldmath $k$}}

\def\pom{{I\!\!P}}

\begin{document}

\title{Testing Non-linear Evolution with Running Coupling Corrections in $ep$ and $pp$ collisions} 
\author{  M.A. Betemps $^{1,2}$, V.P. Gon\c{c}alves $^2$, J. T. de Santana Amaral $^2$}
\affiliation{$^1$ Conjunto Agrot\'ecnico Visconde da Gra\c{c}a (CAVG) \\Universidade Federal de Pelotas,\\
Av. Ildefonso Simes Lopes, 2791\\ CEP 96060-290, Pelotas, RS, Brazil \\
$^2$ High and Medium Energy Group (GAME), \\
Instituto de F\'{\i}sica e Matem\'atica,  Universidade
Federal de Pelotas, 
Caixa Postal 354, CEP 96010-900, Pelotas, RS, Brazil}
\begin{abstract}
The perturbative QCD predicts that the growth of the gluon density at small-$x$ (high energies) should saturate, forming a  Color Glass Condensate (CGC), which is described in mean field approximation by the Balitsky-Kovchegov (BK) equation.  Recently, the next-to-leading order corrections for the BK equation were derived and a global fit of the inclusive $ep$ HERA data was performed, resulting in a parameterization for the forward scattering amplitude. In this paper we compare this parameterization with the predictions of other phenomenological models  and investigate the saturation physics in diffractive deep inelastic electron-proton scattering and in the forward hadron production in $pp$ collisions. Our results demonstrate that the running coupling BK solution is able to describe these observables.

\end{abstract}
\maketitle
\vspace{1cm}

\section{Introduction}

The understanding of the high energy (small $x$) regime of Quantum Chromodynamics (QCD) has been one of the main challenges of this theory,
which has been intensely investigated through high energy collision experiments.
This regime, where one expects to observe the non-linear behavior predicted
by theoretical developments, has been explored in $ep$ collisions at DESY-HERA and $pp/dA$ collisions at BNL-RHIC and, in a near future, in $pp/pA/AA$ collisions at CERN-LHC. In particular, at high energies, the growth of the parton distribution is expected to saturate, forming a  Color Glass Condensate (CGC), whose evolution with energy is described by an infinite hierarchy of coupled equations for the correlators of  Wilson lines \cite{BAL,BAL1,BAL2,BAL3,BAL4,CGC,CGC1,CGC2,CGC3,CGC4,CGC5,CGC6,CGC7}.  
In the mean field approximation, the first equation of this  hierarchy decouples and boils down to a single non-linear integro-differential  equation: the Balitsky-Kovchegov (BK) equation \cite{BAL,KOVCHEGOV,KOVCHEGOV1}.

The BK equation determines, in the large-$N_c$ (the number of colors) limit, the evolution of the two-point correlation function, which corresponds to the  scattering amplitude ${\cal{N}}(x,r,b)$ of a dipole off the CGC, where $r$ is the dipole size and $b$ the impact parameter. This quantity  encodes the information about the hadronic scattering  and then about the non-linear and quantum effects in the hadron wave function (For recent reviews, see e.g. \cite{hdqcd,hdqcd1,hdqcd2,hdqcd3}). In the last years, several groups have constructed phenomenological models which satisfy the asymptotic behaviors of the leading order BK equation in order to fit the HERA and RHIC data \cite{GBW,GBW2,dipolos2,dipolos3,dipolos4,dipolos5,dipolos6,dipolos7,dipolos8,dipolos9,dipolos10,iim,kkt,dhj,dhj1,Goncalves:2006yt,buw}. In general, it is  assumed that the impact parameter dependence of $\cal{N}$ can be factorized as  ${\cal{N}_{F,A}}(x,r,b) = {\cal{N}_{F,A}}(x,r) S(b)$, where
$S(b)$ is the profile function in impact parameter space and ${\cal N_F}$ and
${\cal N_A}$ are the fundamental and adjoint dipole scattering amplitudes, respectively. The latter can be modelled in the coordinate space, through a simple Glauber-like formula, which reads
\beq
{\cal{N}_A}(x,r) = 1 - \exp\left[ -\frac{1}{4} (r^2 Q_s^2)^{\gamma (x,r^2)} \right] \,\,,
\label{ngeral}
\eeq
where $\gamma$ is the anomalous dimension of the target gluon distribution, and the former can be also parameterized as in (\ref{ngeral}), after the replacement $Q_s^2 \rightarrow Q_s^2\, C_F/C_A = 4/9\, Q_s^2$. The main difference among the distinct phenomenological models comes from the behavior predicted for the anomalous dimension, which determines  the  transition from the non-linear to the extended geometric scaling regime, as well as from the extended geometric scaling to the DGLAP regime. In this paper we restrict our analyses to the model proposed in Ref. \cite{buw}, the so called BUW model, which is able to describe  the $ep$ HERA data for the proton structure function and the hadron spectra measured in $pp$ and $dAu$ collisions at RHIC energy \cite{buw,marcos_vic}. It should be contrasted with, for instance, the IIM \cite{iim}  and DHJ \cite{dhj,dhj1} models, which are only able to describe one of these data sets. Another feature of the BUW model which motivates this analysis is that it explicitly satisfies the property of geometric scaling, which is predicted for the solutions of the BK equation in the asymptotic regime of large energies. In the BUW model, the adjoint dipole scattering amplitude is parameterized in momentum space and is given by
\begin{eqnarray}
{\cal{N}_A}(x,p_T)= - \int d^2 r e^{i\vec{p_T}\cdot \vec{r}}\left[1-\exp\left(-\frac{1}{4}(r^2Q_s^2(x))^{\gamma(p_T,x)}\right)\right]\,\,,
\end{eqnarray}
where $\gamma$ is assumed to be a function of $p_T$, rather than $r$, in order to make easier the evaluation of its Fourier transform, and is given by  $\gamma(p_T,x)=\gamma_s+\Delta\gamma(p_T,x)$, where  $\gamma_s = 0.628$ and  \cite{buw}
\begin{eqnarray}
\label{BUWeq}
\Delta \gamma(p_T,x) = \Delta \gamma_{BUW} =(1-\gamma_s)\frac{(\omega^a-1)}{(\omega^a-1)+b}.
\end{eqnarray}
In the expression above, $\omega  \equiv p_T/Q_s(x)$ and the two free parameters $a=2.82$ and $b=168$ are fitted in such a way to describe the RHIC data on hadron production. It is clear, from Eq.(\ref{BUWeq}), that this model satisfies the property of geometric scaling \cite{scaling,marquet,prl,prl1}, since $\Delta\gamma$ depends on $x$ and $p_T$ only through the variable $p_T/Q_s(x)$.
 Besides, in comparison with other phenomenological parameterizations, in the BUW model, the large $p_T$ limit, $\gamma \rightarrow 1$, is approached much faster, which implies different predictions for the large $p_T$ slope of the hadron and photon yield (For a detailed discussion see Ref. \cite{marcos_vic}).

Recently, the next-to-leading order corrections to BK equation were calculated  \cite{kovwei1,kovwei2,javier_kov,balnlo,balnlo2,kovwei3} through the ressumation of $\alpha_s N_f$ contributions to all orders, where $N_f$ is the number of flavors. Such calculation allows one to estimate the soft gluon emission and running coupling corrections to the evolution kernel and, in particular, the authors have verified that  the dominant contributions come from the running coupling corrections, which allows to determine the scale of the running coupling in the kernel. The solution of the improved BK equation was studied in detail in Refs. \cite{javier_kov,javier_prl}. Basically, one has that the running of the coupling reduces the speed of the evolution to values compatible with experimental data, with the geometric scaling regime being reached only at ultra-high energies. In Ref. \cite{javier_prl}, the solution of the improved BK equation was used to calculate the pseudorapidity density of charged particles produced in nucleus-nucleus collisions and a remarkable good agreement with the RHIC data was observed.

More recently, a global analysis of the small $x$ data for the proton structure function using the improved BK equation was performed \cite{bkrunning} (See also Ref. \cite{weigert}). In contrast to the  BK  equation at leading logarithmic $\alpha_s \ln (1/x)$ approximation, which  fails to describe data, the inclusion
of running coupling effects to evolution renders BK equation compatible with them. The impact parameter dependence was not taken into account, the normalization of the dipole cross section was fitted to data and two distinct initial conditions, inspired in the Golec Biernat-Wusthoff (GBW) \cite{GBW} and McLerran-Venugopalan (MV) \cite{MV} models, were considered. The predictions resulted to be almost independent of the initial conditions and, besides, it was observed that it is impossible to describe the experimental data using only the linear limit of the BK equation, which is equivalent to Balitsky-Fadin-Kuraev-Lipatov (BFKL) equation \cite{bfkl}.

In this paper we compare the parameterization proposed in \cite{bkrunning} with the predictions of the BUW model. A basic difference among these models is the set of experimental data used in order to constrain the free parameters of the parameterizations or the initial condition of the running coupling (RC) BK evolution equation. For instance, while the BUW model \cite{buw} uses the RHIC data for the forward hadron production in $dAu$ collisions, in Ref. \cite{bkrunning}  the initial condition is constrained using the small-$x$ HERA data for the proton structure function. In the original papers, the predictions of these two  models were compared with the $\sigma^{\gamma^* p}$ and $F_L$ HERA data, respectively. However, the resulting predictions for  observables measured in diffractive deep inelastic scattering (DDIS) are still an open question. This is one of our goals. Another one is to calculate for the first time the forward hadron production in $pp$ collisions using the solution of the RC BK evolution equation. This allows one to check the generalized dilute-dense factorization used to calculate the observables in hadron-hadron collisions and to verify if it is possible an unified description of the high energy regime in $ep$ and $pp$ collisions  using the Color Glass Condensate formalism.

This paper is organized as follows. In Section II we present the BK
equation in both leading-order and with the improvements provided by the inclusion
of the running of the coupling, and compare the behavior of the resulting
dipole-target scattering amplitude with well-known parametrizations based on the
saturation physics. Section III is devoted to the description of recent HERA data on
the diffractive proton structure function $F_2^{D(3)}$, while in Section IV hadron production in $pp$ collisions at RHIC is investigated. The conclusions are presented in Section V.


\section{BK equation and running coupling effects}

The Balitsky-Kovchegov equation is the simplest non-linear evolution equation
for the dipole-hadron scattering amplitude, being actually a mean field version
of the first equation of the Balitsky hierarchy \cite{BAL,BAL1}. In leading
order (LO), and in the translational invariance approximation---in which the scattering
amplitude does not depend on the collision impact parameter $\bm{b}$---it reads

	\begin{equation}\label{eq:bklo}
		\frac{\partial {\cal{N}}_F(r,Y)}{\partial Y} = \int {\rm d}\bm{r_1}\, K^{\rm{LO}}
		(\bm{r,r_1,r_2})
		[{\cal{N}}_F(r_1,Y)+{\cal{N}}_F(r_2,Y)-{\cal{N}}_F(r,Y)-{\cal{N}}_F(r_1,Y){\cal{N}}_F(r_2,Y)],
	\end{equation}
where ${\cal{N}}_F(r,Y)$ is the (fundamental) scattering amplitude for a dipole (a quark-antiquark pair)
off a target, with transverse size $r\equiv |\bm{r}|$, $Y\equiv \ln(x_0/x)$ ($x_0$ is the value of $x$ where the evolution starts), and $\bm{r_2 = r-r_1}$. $K^{\rm{LO}}$ is the evolution kernel, given by

	\begin{equation}\label{eq:klo}
		K^{\rm{LO}}(\bm{r,r_1,r_2}) = \frac{N_c\alpha_s}{2\pi^2}\frac{r^2}{r_1^2r_2^2},
	\end{equation}
where $\alpha_s$ is the (fixed) strong coupling constant. This equation is a
generalization of the linear BFKL equation (which corresponds of the first three terms), with the inclusion
of the (non-linear) quadratic term, which damps the indefinite growth of the amplitude
with energy predicted by BFKL evolution. It has been shown \cite{mp} to be in the same
universality class of the Fisher-Kolmogorov-Pertovsky-Piscounov (FKPP) equation
\cite{fkpp} and, as a consequence, it admits the so-called traveling wave solutions.
This means that, at asymptotic rapidities, the scattering amplitude is a wavefront which
travels to larger values of $r$ as $Y$ increases, keeping its shape unchanged. Thus,
in such asymptotic regime, instead of depending separately on $r$ and $Y$, the amplitude
depends on the combined variable $rQ_s(Y)$, where $Q_s(Y)$ is the saturation scale. This property of the solution of BK equation is a natural
explanation to the {\it geometric scaling}, a phenomenological feature observed at the
DESY $ep$ collider HERA, in the measurements of inclusive and exclusive processes
\cite{scaling,marquet,prl,prl1}.

Although having its properties been intensely studied and understood, both numerically
and analytically, the LO BK equation presents some difficulties when applied to study
DIS small-$x$ data. In particular, some studies concerning this equation
\cite{IANCUGEO,MT02,AB01,BRAUN03,AAMS05} have
shown that the resulting saturation scale grows much faster with increasing energy
($Q_s^2\sim x^{-\lambda}$, with $\lambda\simeq 4.88N_c\alpha_s/\pi \approx 0.5$ for  $\alpha_s = 0.2$) than that
extracted from phenomenology ($\lambda \sim 0.2-0.3$). This difficulty could be solved by
considering smaller values of the strong coupling constant $\alpha_s$, but this procedure
would lead to physically unrealistic values. One can conclude that higher order corrections
to LO BK equation should be taken into account to make it able to describe the available
small-$x$ data.

The calculation of the running coupling corrections to BK evolution kernel was explicitly
performed in \cite{kovwei1,kovwei2,balnlo,balnlo2,kovwei3}, where the authors included $\alpha_sN_f$ corrections to the kernel to all orders. The  improved  BK equation is given in terms of a  
running coupling and a subtraction term, with the latter accounting for conformal, non running coupling contributions. In the prescription proposed by Balitsky in \cite{balnlo} to single out the ultra-violet divergent contributions from the finite ones that originate after the resummation of quark loops, the contribution of the subtraction term is minimized at large energies. In \cite{bkrunning} this contribution was disregarded, and the improved BK equation was numerically solved replacing the leading order kernel  in Eq. (\ref{eq:bklo}) by the modified kernel which includes the running coupling
corrections and  is given by \cite{balnlo}
	\begin{equation}\label{eq:krun}
		K^{\rm{Bal}}(\bm{r,r_1,r_2})=\frac{N_c\alpha_s(r^2)}{2\pi^2}
		\left[\frac{r^2}{r_1^2r_2^2} + \frac{1}{r_1^2}\left(\frac{\alpha_s(r_1^2)}
		{\alpha_s(r_2^2)}-1\right)+\frac{1}{r_2^2}\left(\frac{\alpha_s(r_2^2)}
		{\alpha_s(r_1^2)}-1\right)\right].
	\end{equation}
From a recent numerical study of the improved BK equation \cite{javier_kov}, it has been confirmed that the running coupling corrections lead to a considerable
increase in the anomalous dimension and to a slow-down of the evolution
speed, which implies, for example, a slower growth of the saturation scale with
energy, in contrast with the faster growth predicted by the LO BK equation.
As it was discussed in the Introduction, the improved BK equation has been shown to
be really successful when applied to the description of the $ep$ HERA data for the proton structure function, which motivates us to extend the study for other observables.


\begin{figure}[t]
\centerline{
{\psfig{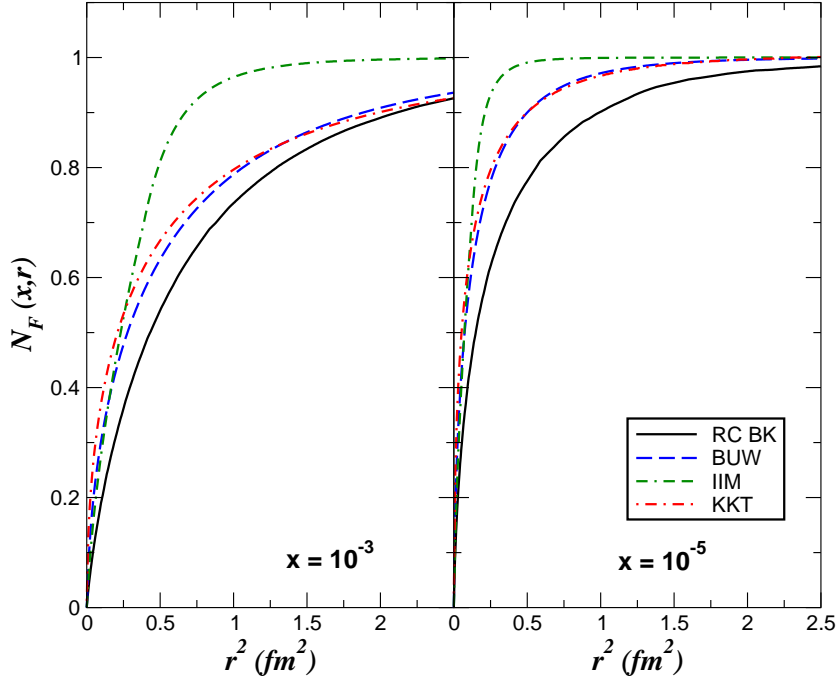}}}
\caption{Dependence of the fundamental dipole scattering amplitude  in the squared pair separation $\rr^2$ at 
different values of $x$.}
\label{fig1}
\end{figure}

\subsection{BK with running coupling and phenomenological models}

Before performing our phenomenological study using the improved BK equation,
whose solution we will call RC BK from now on, it is interesting, at this point, to investigate the behavior of its solution and compare it with those from the
phenomenological models based on saturation physics, which are able to describe HERA and/or RHIC data. It is important to point out that in order to make this comparison,
as well as the phenomenological study to be developed in the following sections, we make
use of the public-use code available in \cite{code}.

In Fig.\ref{fig1} we show the pair separation dependence of the fundamental scattering amplitude $\cal{N}_F$ for different values of $x$ (For a related discussion see Ref. \cite{testing}). From this figure one can observe that while the KKT \cite{kkt} and BUW parameterizations present a similar behavior for small $\rr^2$, the RC BK one predicts a smoother dependence. In the limit of large pair separations, or large dipoles, the IIM \cite{iim} parameterization saturates, while the BUW, KKT and RC BK ones still present a residual dependence, demonstrating that for these models, the asymptotic regime is only reached for very large pair separations. A characteristic feature which is evident in the  IIM model is that the dipole scattering amplitude saturates for smaller dipoles when $x$ assumes smaller values. An important aspect to be emphasized is the large difference between the predictions in the transition region, which is expected to be probed at  HERA and RHIC.


\begin{figure}[t]
\centerline{
{\psfig{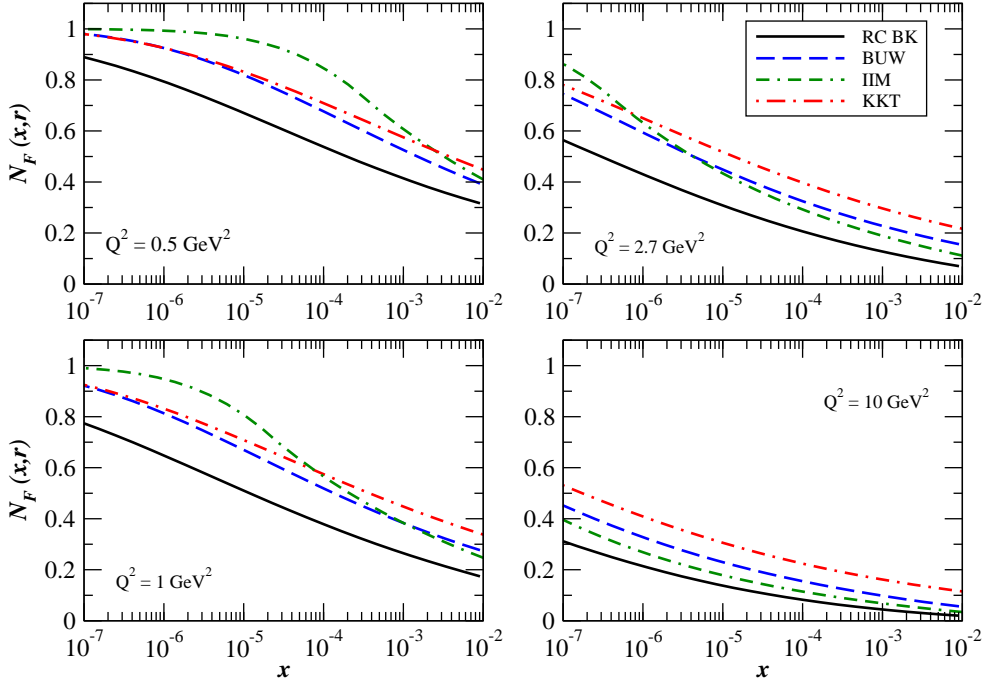}}}
\caption{Energy dependence of the fundamental scattering amplitude as a function of $x$ at different photon virtualities.}
\label{fig2}
\end{figure}

A basic feature of the phenomenological  parameterizations is that they grow in the region of small values of $\rr Q_s$ as a power of $\rr Q_s$, i.e. $ {\cal{N}}(\rr,x) \propto (\rr Q_s)^{2 \gamma_{eff}}$. However,   $\gamma_{eff}$ is different in each model, being $\gamma \le 1$ for the IIM model and about 
$\frac{1}{2}$ for the KKT and BUW ones. This implies a different $\rr Q_s$ dependence of the dipole scattering amplitudes and dipole cross sections.  Since the saturation scale drives the energy dependence of the dipole cross section, these models present a very distinct energy dependence. This can clearly be seen  in Fig. \ref{fig2}, where we present the $x$ dependence of the dipole scattering amplitudes  for different values of the squared pair separation given by $\rr^2 = 1/Q^2$. We observe that for large $Q^2$ (small pair separation) the dipole scattering amplitude is dominated by the linear limit. Since the models have different behaviors in this limit, the energy dependence is also different, with the IIM model presenting the strongest growth at small $x$.  On the other hand, the KKT and BUW models predict the two smallest growths with the energy. 
At large pair separations $r > 1/Q_s$, which characterize the saturation regime, the IIM model predicts the saturation of the dipole scattering amplitude, while the KKT and BUW ones still present a growth at small values $x$.  In comparison,  the RC BK solution predicts a smooth growth, with a smaller normalization and a delayed saturation of the fundamental scattering amplitude at small values of $x$. Basically, the asymptotic saturation regime is only observed for very small values of $x$, beyond the kinematical range of HERA.



\section{Diffractive DIS in the CGC Formalism}

In the last years, significant progress in understanding diffraction has been made at the $ep$ collider HERA (See, e.g. Refs. \cite{MW,H,PREDAZZI}). Currently, there exist many attempts to describe the diffractive part of 
the deep inelastic cross section within  pQCD (See, e.g. Refs. \cite{GBW,fss,MRW,Brod}). One of the most successful approaches is the saturation one \cite{GBW2,dipolos9,fss}  based  on the dipole picture of DIS \cite{dipole,dipole2}. It naturally incorporates the description of both inclusive and diffractive events in a common theoretical framework, as the 
same dipole scattering amplitude enters in the formulation of the inclusive and diffractive cross sections. In the studies of saturation effects in DDIS,  non-linear evolution equations for the dipole scattering amplitude have been derived \cite{KL,KW,W}, new measurements proposed \cite{slopedif1,slopedif2,MS,GM}  and the charm contribution estimated \cite{charmdif,golec}. As shown in Ref. \cite{GBW2}, the total diffractive cross section is much more sensitive to large-size dipoles than the inclusive one. Saturation effects  screen large-size dipole (soft) contributions, so that a fairly large fraction of the cross section is hard and hence eligible for a perturbative treatment. Therefore,  the study of diffractive processes 
becomes fundamental in order to constrain the QCD dynamics at high energies.

The diffractive process can be analyzed in detail studying the behavior of the diffractive structure function $F_2^{D (3)}(Q^{2}, \beta, x_{I\!\!P})$. In Refs. \cite{GBW2,dipole} the authors have derived expressions for $F_2^{D (3)}$ directly in the transverse momentum space and then transformed to impact parameter space where the dipole approach can be applied. Following Ref. \cite{GBW2} we assume that the diffractive structure function is given by
\begin{equation}
F_2^{D (3)} (Q^{2}, \beta, x_{I\!\!P}) = F^{D}_{q\bar{q},L} + F^{D}_{q\bar{q},T} + F^{D}_{q\bar{q}g,T}
\label{soma}
\end{equation}
where $T$ and $L$ refer to the polarization of the virtual photon. For the $q\bar{q}g$ contribution only the transverse polarization is considered, since the longitudinal counterpart has no leading logarithm in $Q^2$.  The computation of the different contributions was done in Refs. \cite{wusthoff,GBW2,dipole} and here we quote only the final results:
\begin{equation}
  x_{I\!\!P}F^{D}_{q\bar{q},L}(Q^{2}, \beta, x_{I\!\!P})=
\frac{3 Q^{6}}{32 \pi^{4} \beta B_D} \sum_{f} e_{f}^{2} 
 2\int_{\alpha_{0}}^{1/2} d\alpha \alpha^{3}(1-\alpha)^{3} \Phi_{0},
\label{qqbl}
\end{equation}
\begin{equation}
 x_{I\!\!P}F^{D}_{q\bar{q},T}(Q^{2}, \beta, x_{I\!\!P}) =  
 \frac{3 Q^{4}}{128\pi^{4} \beta B_D}  \sum_{f} e_{f}^{2} 
 2\int_{\alpha_{0}}^{1/2} d\alpha \alpha(1-\alpha) 
\left\{ \epsilon^{2}[\alpha^{2} + (1-\alpha)^{2}] \Phi_{1} + m_f^{2} \Phi_{0}  \right\}   
\label{qqbt}
\end{equation}
where the lower limit of the integral over $\alpha$ is given by $\alpha_{0} = \frac{1}{2} \, \left(1 - \sqrt{1 - \frac{4m_{f}^{2}}{M_X^{2}}}\right)
$ and we have introduced the auxiliary functions \cite{fss}:
\begin{equation}
\Phi_{0,1}  \equiv  \left(\int_{0}^{\infty}r dr K_{0 ,1}(\epsilon r)\sigma_{dip}(x_{I\!\!P},r) J_{0 ,1}(kr) \right)^2.
\label{fi}
\end{equation}
Following Refs.  \cite{wusthoff,GBW2,nikqqg,golec}, here we calculate the $q\bar{q}g$ contribution within the dipole picture at leading $\ln Q^2$ accuracy, where it reads
 \begin{eqnarray}
   \lefteqn{x_{I\!\!P}F^{D}_{q\bar{q}g,T}(Q^{2}, \beta, x_{I\!\!P}) 
  =  \frac{81 \beta \alpha_{S} }{512 \pi^{5} B_D} \sum_{f} e_{f}^{2} 
 \int_{\beta}^{1}\frac{\mbox{d}z}{(1 - z)^{3}} 
 \left[ \left(1- \frac{\beta}{z}\right)^{2} +  \left(\frac{\beta}{z}\right)^{2} \right] } \label{qqg} \\
  & \times & \int_{0}^{(1-z)Q^{2}}\mbox{d} k_{t}^{2} \ln \left(\frac{(1-z)Q^{2}}{k_{t}^{2}}\right) 
\left[ \int_{0}^{\infty} u \mbox{d}u \; \sigma_{dip}(u / k_{t}, x_{I\!\!P}) 
   K_{2}\left( \sqrt{\frac{z}{1-z} u^{2}}\right)  J_{2}(u) \right]^{2}.\nonumber
\end{eqnarray} 
As pointed in Ref. \cite{dipolos9}, at small $\beta$ and low $Q^2$, the leading $\ln (1/\beta)$ terms should be resumed and the above expression should be modified. However, as a description with the same quality using the Eq. (\ref{qqg}) is possible by adjusting the coupling \cite{dipolos9}, in what follows we will use this expression for our phenomenological studies. 
We  use the standard notation for the variables $\beta = Q^2 / (M_X^2 + Q^2)$, $
x_{I\!\!P} = (M_X^2 + Q^2)/(W^2 + Q^2)$ and $x = Q^2/(W^2 + Q^2) = \beta x_{\pom}$, 
where $M_X$ is the invariant mass of the diffractive system, $B_D$ is the diffractive slope and $W$ the total energy of the 
$\gamma ^* p$ system.




The dipole cross section, $\sigma_{dip} (x, \rr)$, is determined by the QCD dynamics, being closely related to the solution of the QCD non-linear evolution equations (For recent reviews see, e.g. Refs. \cite{hdqcd,hdqcd1,hdqcd2,hdqcd3})
\begin{eqnarray}
\sigma_{dip} (x,\rr)=2 \int d^2 \rb \, {\cal{N}_F}(x,\rr,\rb)\,\,,
\end{eqnarray}
where ${\cal{N}_F}$ is the  fundamental dipole-target forward scattering amplitude for a given impact parameter $\rb$, which encodes all the
information about the hadronic scattering, and thus about the
non-linear and quantum effects in the hadron wave function. In what follows we estimate, for the first time, the diffractive structure function considering the BUW and RC BK parameterizations as input in the calculations. Both consider an independence of the amplitude with respect to the impact parameter, which implies that $\sigma_{dip} = \sigma_0 \, {\cal{N}}(x,\rr)$, $\sigma_0$ being a free parameter fitted to { the respective} data. It is important to emphasize that the value of  $\sigma_0$ directly constrains the diffractive slope $B_D$, since both are related by $\sigma_0 = 4\pi B_D$ if we assume a Gaussian form factor for the proton \cite{dipolos9}. 

\begin{figure}[t]
\centerline{
{\psfig{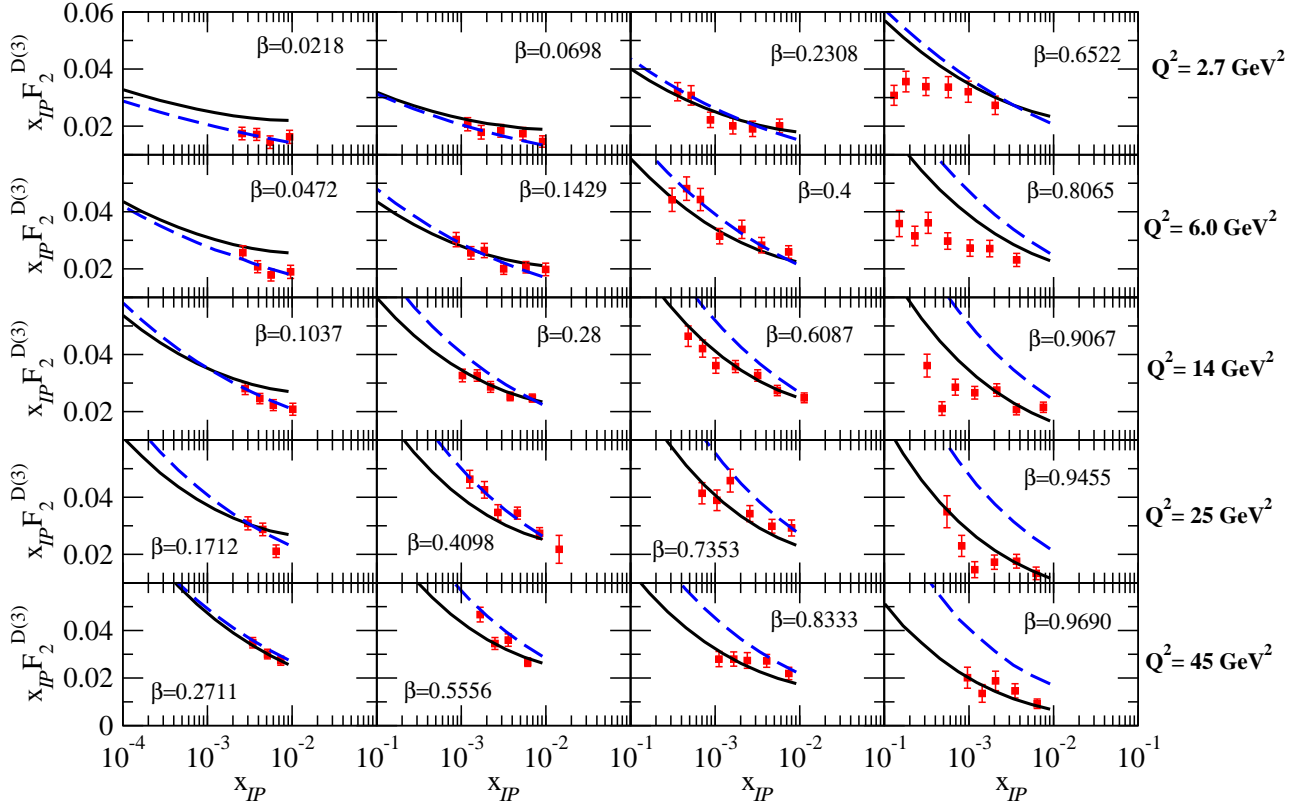}}}
\caption{Predictions for $F_2^{D (3)} (Q^{2}, \beta, x_{I\!\!P})$  compared with the ZEUS data \cite{zeusdata}. Solid line: RC BK model; Dashed line: BUW model.}
\label{fig3}
\end{figure}

In order to calculate the diffractive structure function and compare with the HERA data we need to specify the diffractive slope $B_D$ and the coupling $\alpha_s$, which determine the normalization of $F_2^{D (3)} (Q^{2}, \beta, x_{I\!\!P})$. In particular, the magnitude of the $q\bar{q}g$ contribution is strongly dependent on the value of $\alpha_s$.
The BUW model implies $B_D = 4.3$ GeV$^{-2}$, while the RC BK one implies $B_D = 6.7$ GeV$^{-2}$. Both values are in reasonable agreement with the experimental data \cite{aktas}. On the other hand, we are still free to choose the value of $\alpha_s$. Following \cite{fss}, we assume $\alpha_s = 0.15$. In a more detailed study we could consider its running with $Q^2$ or perform a fit to experimental data. However, as our goal is to check if these models can describe the experimental data, we postpone this study to a future publication. 

In which concerns the RC BK model, it is important to specify the initial
conditions for the evolution of the scattering amplitude. In \cite{bkrunning},
two families of initial conditions were considered, one inspired in the GBW
saturation model \cite{GBW}, and the another one in the MV model \cite{MV}.
In the present analysis, we use the latter, given by
	\begin{equation}
		\calN^{\rm{MV}}(r,Y=0)=1-\exp\left[-\left(\frac{r^2Q_{s0}^2}{4}
		\right)^\gamma\ln\left(\frac{1}{r\Lambda_{\rm{QCD}}}+e\right) \right],
	\end{equation}
where $Q_{s0}^2$ is the initial saturation scale squared and $\gamma$ is an anomalous dimension. Both parameters are obtained from the fit to $F_2$ data and are given by 
$Q_{s0}^2=0.15$ GeV$^2$ and $\gamma = 1.13$.


The diffractive cross section $ep \rightarrow eXY$ have been measured by the H1 and ZEUS experiments at HERA tagging the proton in the final state ($Y = p$) or selecting events with a large rapidity gap between the systems $X$ and $Y$ in the case of H1  and using the $M_X$-method in case of ZEUS. The distinct methods and experimental cuts used by the H1 and ZEUS collaborations imply a difference of normalization between their data. Moreover, while the ZEUS data are given for the diffractive structure function $F_2^{D (3)}$, the H1 ones are presented for the reduced cross section which is expressed in terms of a combination of diffractive structure functions and kinematical factors. As our predictions are for $F_2^{D(3)}$, as given in Eq. (\ref{soma}), we restrict our comparison to the recent  ZEUS data \cite{zeusdata}. Furthermore, as the dipole model is more suitable for the description of the diffractive structure functions in the region of low and moderate $Q^2$, we restrict our comparison to the experimental data in the kinematical region of $Q^2 < 50$ GeV$^2$ and $x_{I\!\!P} \le 10^{-2}$.
In Fig. \ref{fig3} we compare the predictions of the BUW and RC BK models with the
ZEUS data \cite{zeusdata} for five values of the photon virtuality $Q^2$. 
We have that
both models describe reasonably the experimental data at medium values of $\beta$.
However, the predictions differ at large and small  $\beta$. In particular, the
BUW model over predicts the data  at large $\beta$ and large $Q^2$, while the RC
BK model describes quite well this same data set. This result can be associated
to the features of the BUW model, which assumes the geometric scaling property
and predicts that the limit $\gamma \rightarrow 1$ is rapidly approached. As a
consequence, one obtains different predictions for the momentum dependence of
 the dipole cross section in comparison with, for instance, the IIM model,
which is able to describe the data in this  kinematical range \cite{fss}. It is important to emphasize that in our study the slope parameter is not a free parameter, being derived from $\sigma_0$. An alternative is to assume $B_D$ as a free parameter to be obtained from the fit. However, this procedure would change the normalization  of the BUW  predictions by a constant factor for all values of $\beta$ and $Q^2$. A larger value of $B_D$ would improve the description at large $\beta$, but would lead to an underprediction of the small $\beta$ data. Consequently, the conclusion that the BUW model is only able to describe a restricted kinematical range would not be modified. 
In the case of the RC BK model,  it  over predicts
the diffractive structure function at small $\beta$ and $Q^2$. Although it is possible to get rid of
this over prediction by modifying the value of the coupling $\alpha_s$, this
would lead to non-physical small values for it. This may be an indication
that a generalized expression for $F_2^{D (3)} (Q^{2}, \beta, x_{I\!\!P})$,
as that proposed in \cite{dipolos9}, should be used to calculate the
diffractive structure function at small-$\beta$. 
 
\section{Forward hadron production} 

Another source of information on QCD dynamics at high parton density is the forward hadron production in hadron-hadron collisions at RHIC. In particular, the observed suppression of the normalized hadron production transverse momentum in $dAu$ collisions as compared to $pp$ collisions has been considered an important signature of the Color Glass Condensate physics (For a review see e.g. \cite{hdqcd2}). As pointed in Ref. \cite{difusivo}, the forward hadron production in hadron-hadron collisions is  a typical example of a dilute-dense process, which is an ideal system to study the small-$x$ components of the  target wave function.  In this case the cross section is expressed as a convolution of the standard parton distributions for the dilute projectile, the dipole-hadron scattering amplitude (which includes the high-density effects) and the parton fragmentation functions.  Basically, assuming this generalized dense-dilute factorization, the minimum bias invariant yield for single-inclusive hadron production in hadron-hadron processes is described in the CGC formalism  by \cite{dhj,dhj1,arata}
\begin{eqnarray}\label{eq:hadron}
 \frac{d^2N^{pp(A)\rightarrow hX}} {dyd^2p_T}&=& \frac{1}{(2\pi)^2}
\int_{x_F}^1 dx_1 \frac{x_1}{x_F}\left[ f_{q/p}(x_1,p_T^2){\cal{N}_F}\left(x_2,\frac{x_1}{x_F}p_T\right)
D_{h/q}\left(\frac{x_F}{x_1},p_T^2\right)
\right.\nonumber\\
&+& \left.   f_{g/p}(x_1,p_T^2){\cal{N}_A}\left(x_2,\frac{x_1}{x_F}p_T\right)D_{h/g}\left(\frac{x_F}{x_1},p_T^2\right)  \right]\,\,,
\label{eq:final}
\end{eqnarray}
where $p_T$, $y$ and $x_F$ are the transverse momentum, rapidity and the Feynman-$x$
of the produced hadron, respectively. The variable $x_1$ denotes the momentum
fraction of a projectile parton,    $f(x_1,p_T^2)$ is the projectile parton
distribution functions  and $D(z, p_T^2)$ the parton fragmentation
functions into hadrons. These quantities  evolve according to the 
DGLAP evolution equations \cite{dglap,dglap1,dglap2} and obey the momentum
sum-rule. In Eq. (\ref{eq:final}), ${\cal{N}_F}(x,\rk)$  and  ${\cal{N}_A} (x,\rk)$ are the fundamental and adjoint representations of the forward dipole amplitude in momentum space, which represents the probability for scattering of a quark and a gluon off the nucleus, respectively.  Moreover, $x_F=\frac{p_T}{\sqrt{s}}e^{y}$ and the momentum fraction of the target partons is given by $x_2=x_1e^{-2y}$ (For details see e.g. \cite{arata}).

\begin{figure}[t]
\centerline{
{\psfig{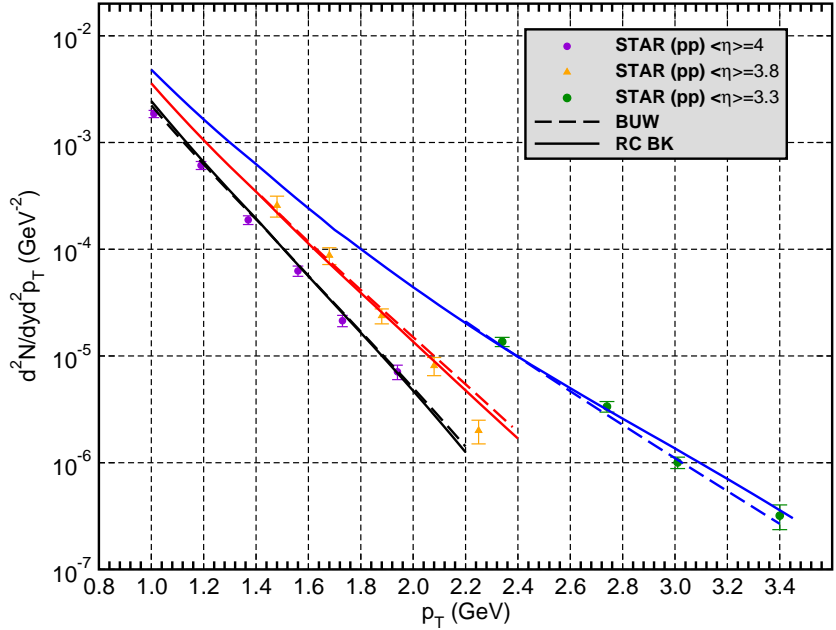}}}
\caption{Inclusive $\pi^0$ production cross section in $pp$ collisions at RHIC energies. Data from STAR collaboration \cite{star}. We assume $K (\eta = 4.0) = K (\eta = 3.8) = 1.4$ and $K (\eta = 3.3) = 1.0$ for the RC BK (solid line)  and BUW (long-dashed line) predictions.}
\label{fig4}
\end{figure}

In the last years, several models have been proposed to describe the hadron spectra in $dAu$ collisions, obtaining a satisfactory description of these experimental data \cite{kkt,dhj,dhj1,Goncalves:2006yt,buw}. In general, these models have been extended to $pp$ collisions  in order to calculate the ratio $R_{hA}$ without a comparison with the corresponding experimental data. In Ref. \cite{marcos_vic} the authors estimated the hadron production in $pp$ and $dAu$ collisions in a same theoretical formalism and compared these predictions with the experimental data. They fixed the only free parameter in the calculations (the $K$-factor) and obtained parameter-free predictions for the nuclear modification ratio $R_{hA}$. The comparison with the experimental data demonstrated that the BUW model, which assumes the geometric scaling property, is adequate for the RHIC kinematical range.  Furthermore, for the first time, it is possible to estimate the hadron spectra in $pp$ collisions using as input the solution of the BK evolution equation. In what follows  we calculate the spectra and compare the BUW predictions with those from the RC BK equation. 

For the initial conditions, we use the MV-inspired ones, as it was done in the
case of diffraction. In Eq.(\ref{eq:hadron}), both fundamental and adjoint
scattering amplitudes enters in the expression, but only the fundamental one is
described by BK equation. Thus, we must perform the transformation from $\calN_F$
to $\calN_A$ described in the Introduction, but now at the level of the $x$ variable,
and not the squared saturation scale, since we have been working directly with the solution of BK equation. More specifically, from the transformation
$Q_s^2 \rightarrow Q_s^2\, C_F/C_A = 4/9\, Q_s^2$, it is straightforward to find
the transformation for $x$ through the relation $Q_{s}^2 = (x_0/x)^\lambda$.

In Fig. \ref{fig4} we present our predictions for the production of forward $\pi^0$ mesons in $pp$ collisions at forward rapidities and compare our results with the STAR data \cite{star}. 
In our calculations we use the CTEQ5L parameterization \cite{cteq}  for the parton distribution functions  and the KKP parameterization for the  fragmentation functions \cite{kkp}. As in previous calculations \cite{dhj,dhj1,Goncalves:2006yt,buw,marcos_vic} there is only one free parameter in our calculation: the $K$-factor. It is determined in such way to provide the best description of the experimental data and is fixed for each rapidity. We can see that both models describe quite well the experimental data assuming an identical $K$-factor. This result is directly associated to the similar behavior of the dipole scattering amplitude at large pair separations predicted by these two models (See Fig. \ref{fig1}), which is the region probed in $pp$ collisions at forward rapidities.

\section{Conclusions}

 Recent calculations of the running coupling corrections to the BK equation allowed a global fit to the proton structure function to be performed within an unprecedented level of precision. This result motivates the study of other observables in order to test the non-linear small-$x$ evolution. In this paper we have extended the predictions of the running coupling BK equation to diffractive observables in $ep$ collisions and to the forward hadron  production in $pp$ collisions. Our results demonstrate that the current experimental data can be described using the solution of the BK equation. Consequently, we believe that this framework is adequate to calculate the observables which would be studied in the future colliders.



\begin{acknowledgments} This work was  partially 
financed by the Brazilian funding
agencies CNPq and FAPERGS.
\end{acknowledgments}




\begin{thebibliography}{99}







\bibitem{BAL}  I. I. Balitsky,   Nucl. Phys. {\bf  B463} (1996) 99.
\bibitem{BAL1}  I. I. Balitsky, Phys. Rev. Lett. {\bf 81} (1998) 2024.  

\bibitem{BAL2} I. I. Balitsky, Phys. Rev. D  {\bf 60} (1999) 014020.
\bibitem{BAL3} I. I. Balitsky,   Phys. Lett. B  {\bf 518} (2001) 235.
\bibitem{BAL4}  I.I. Balitsky and  A.V. Belitsky, Nucl. Phys. {\bf B629}  (2002) 290. 
 
\bibitem{CGC} E. Iancu, A. Leonidov and L. McLerran, Nucl.Phys.  
{\bf A692} (2001) 583.
\bibitem{CGC1}  E. Ferreiro, E. Iancu, A. Leonidov and  L. McLerran, Nucl. Phys. {\bf A701} (2002) 489.
\bibitem{CGC2} J. Jalilian-Marian, A. Kovner, L. McLerran  and  H. 
Weigert, Phys. Rev. D {\bf 55}  (1997) 5414.
\bibitem{CGC3} J. Jalilian-Marian, A. Kovner and  H. 
Weigert, Phys. Rev. D {\bf 59}  (1999) 014014.
\bibitem{CGC4}  J. Jalilian-Marian, A. Kovner and  H. 
Weigert, Phys. Rev. D{\bf 59}  (1999) 014015.
\bibitem{CGC5} J. Jalilian-Marian, A. Kovner and  H. 
Weigert, Phys. Rev. D {\bf 59} (1999)  034007.   
\bibitem{CGC6} A. Kovner, J. Guilherme Milhano and  H. Weigert,   Phys. Rev. D {\bf 62} 
(2000)  114005.
\bibitem{CGC7}  H. Weigert, Nucl. Phys.  {\bf A703} (2002)  823. 
 
 
\bibitem{KOVCHEGOV}  
Y.V. Kovchegov,  Phys. Rev. D {\bf 60},  034008 (1999).
\bibitem{KOVCHEGOV1} Y.V. Kovchegov,  Phys. Rev. D {\bf 61} 074018 (2000). 



\bibitem{hdqcd}
  E.~Iancu and R.~Venugopalan,
  arXiv:hep-ph/0303204.
\bibitem{hdqcd1}  
  H.~Weigert,
  Prog.\ Part.\ Nucl.\ Phys.\  {\bf 55}(2005) 461.
\bibitem{hdqcd2} 
   J.~Jalilian-Marian and Y.~V.~Kovchegov,
  Prog.\ Part.\ Nucl.\ Phys.\  {\bf 56} (2006) 104.

\bibitem{hdqcd3}
  E.~Iancu,
  Nucl.\ Phys.\ Proc.\ Suppl.\  {\bf 191} (2009) 281



\bibitem{GBW} K. Golec-Biernat and  M. Wusthoff,  Phys. Rev. D {\bf 59} (1999) 014017.
\bibitem{GBW2}  K. Golec-Biernat and  M. Wusthoff,  Phys. Rev. D {\bf D60}  (1999) 114023.
\bibitem{dipolos2} J. Bartels, K. Golec-Biernat, H. Kowalski,  Phys. Rev. D {\bf 66} (2002)  014001.
\bibitem{dipolos3} H.~Kowalski and D.~Teaney,
  Phys.\ Rev.\ D {\bf 68} (2003)  114005.
  \bibitem{dipolos4}  H.~Kowalski, L.~Motyka and G.~Watt, Phys.\ Rev.\  D {\bf 74}(2006)  074016.
  \bibitem{dipolos5}  K.~Golec-Biernat and S.~Sapeta,
  Phys.\ Rev.\ D {\bf 74}  (2006) 054032.
\bibitem{dipolos6} J.~T.~de Santana Amaral, M.~B.~Gay Ducati, M.~A.~Betemps and G.~Soyez, Phys.\ Rev.\  D {\bf 76} (2007) 094018. 
\bibitem{dipolos7} C.~Marquet, R.~Peschanski and G.~Soyez, Phys.\ Rev.\  D {\bf 76} (2007) 034011. 

\bibitem{dipolos8} G.~Soyez, Phys.\ Lett.\  B {\bf 655} (2007) 32.

\bibitem{dipolos9}
C. Marquet,  Phys.\ Rev.\  D {\bf 76} (2007) 094017


\bibitem{dipolos10}
G.~Watt and H.~Kowalski,
  Phys.\ Rev.\  D {\bf 78} (2008) 014016


\bibitem{iim} E. Iancu, K. Itakura, S. Munier, 
                  { Phys. Lett.} {\bf B590}  (2004) 199.


\bibitem{kkt} D. Kharzeev, Y.V. Kovchegov and K. Tuchin, 
              { Phys. Lett.} {\bf  B599}  (2004) 23.


\bibitem{dhj}
  A.~Dumitru, A.~Hayashigaki and J.~Jalilian-Marian,
  Nucl.\ Phys.\ A {\bf 765}  (2006) 464.

  \bibitem{dhj1}
  A.~Dumitru, A.~Hayashigaki and J.~Jalilian-Marian, Nucl.\ Phys.\ A {\bf 770}  (2006) 57.

\bibitem{Goncalves:2006yt}
  V.~P.~Goncalves, M.~S.~Kugeratski, M.~V.~T.~Machado and F.~S.~Navarra,
  Phys.\ Lett.\  B {\bf 643}  (2006) 273.
  
\bibitem{buw} D. Boer, A. Utermann, E. Wessels,
Phys.\ Rev.\  D {\bf 77} (2008) 054014.





  

\bibitem{scaling} A. M.  Sta\'sto, K. Golec-Biernat and J. Kwieci\'nski,  Phys. Rev. Lett. 
{\bf 86}  (2001) 596.  


\bibitem{marquet}
C. Marquet and L. Schoeffel,   { Phys. Lett.} {\bf  B639}  (2006) 471.


\bibitem{prl}
  V.~P.~Goncalves and M.~V.~T.~Machado,
  Phys.\ Rev.\ Lett.\  {\bf 91}  (2003) 202002.
  \bibitem{prl1}
  V.~P.~Goncalves and M.~V.~T.~Machado,
 JHEP {\bf 0704}(2007)  028.

\bibitem{marcos_vic}
  M.~A.~Betemps and V.~P.~Goncalves,
  JHEP {\bf 0809} (2008) 019



\bibitem{kovwei1}
  Y.~V.~Kovchegov and H.~Weigert,
  Nucl.\ Phys.\  A {\bf 784} (2007) 188

\bibitem{kovwei2}
  Y.~V.~Kovchegov and H.~Weigert,
  Nucl.\ Phys.\  A {\bf 789} (2007) 260



\bibitem{javier_kov}
  J.~L.~Albacete and Y.~V.~Kovchegov,
  Phys.\ Rev.\  D {\bf 75} (2007) 125021

\bibitem{balnlo}
  I.~Balitsky,
  Phys.\ Rev.\  D {\bf 75} (2007) 014001

\bibitem{balnlo2}
  I.~Balitsky and G.~A.~Chirilli,
  Phys.\ Rev.\  D {\bf 77} (2008) 014019

\bibitem{kovwei3}
  Y.~V.~Kovchegov, J.~Kuokkanen, K.~Rummukainen and H.~Weigert,
  Nucl.\ Phys.\  A {\bf 823} (2009) 47




\bibitem{javier_prl}
  J.~L.~Albacete,
  Phys.\ Rev.\ Lett.\  {\bf 99} (2007) 262301


\bibitem{bkrunning}
  J.~L.~Albacete, N.~Armesto, J.~G.~Milhano and C.~A.~Salgado,
  Phys. Rev. {\bf D80}, 034031 (2009).

\bibitem{IANCUGEO} E.~Iancu, K.~Itakura and  L.~McLerran, Nucl. Phys {\bf A708}, 327 (2002).


\bibitem{MT02} A. H. Mueller and D. N. Triantafyllopoulos, Nucl. Phys.
{\bf B640}, 331 (2002).

\bibitem{AB01} N. Armesto and M. A. Braun, Eur. Phys. J. {\bf C20}, 517
(2001).
\bibitem{BRAUN03} M. A. Braun, Phys. Lett. B 576, 115 (2003).

\bibitem{AAMS05} J. L. Albacete, N. Armesto, J. G. Milhano, C. A. Salgado,
and U. A. Wiedemann, Phys. Rev. D 71, 014003 (2005).

\bibitem{weigert}
  H.~Weigert, J.~Kuokkanen and K.~Rummukainen,
  AIP Conf.\ Proc.\  {\bf 1105} (2009) 394.


\bibitem{MV}
L. McLerran and R. Venugopalan,  Phys.\ Rev.\ D {\bf 49}  (1994) 2233

\bibitem{bfkl}
 L. N. Lipatov, Sov. J. Nucl.
 Phys. \textbf{23}, 338 (1976);
 E. A. Kuraev, L. N. Lipatov and V. S. Fadin,
 Sov. Phys. JETP \textbf{45}, 199 (1977);
 I. I. Balitsky and L. N. Lipatov,
 Sov. J. Nucl. Phys. \textbf{28}, 822 (1978).

 \bibitem{mp}
 S.~Munier and R.~Peschanski,
Phys.\ Rev.\ Lett.\  {\bf 91}, 232001 (2003);
 Phys.\ Rev.\ {\bf D69}, 034008 (2004);
 Phys.\ Rev.\ {\bf D70}, 077503 (2004).

 \bibitem{fkpp}
 R. A. Fisher, Ann. Eugenics \textbf{7}, 355 (1937);
 A. Kolmogorov, I. Petrovsky and N. Piscounov,
 Moscou Univ. Bull. Math. \textbf{A1}, 1 (1937).

\bibitem{code} http://www-fp.usc.es/phenom/rcbk

\bibitem{testing}
  M.~S.~Kugeratski, V.~P.~Goncalves and F.~S.~Navarra,
  Eur.\ Phys.\ J.\  C {\bf 44} (2005) 577







\bibitem{MW}
  M.~Wusthoff and A.~D.~Martin,
  J.\ Phys.\ G {\bf 25} (1999) R309

\bibitem{H}
  A.~Hebecker,
  Phys.\ Rept.\  {\bf 331} (2000) 1


\bibitem{PREDAZZI}  V.~Barone and E.~Predazzi, \textit{High-Energy Particle Diffraction}, 
Springer-Verlag, Berlin Heidelberg, (2002).
  
   


\bibitem{fss} J. R. Forshaw, R. Sandapen and G. Shaw,  Phys. Lett. 
               \textbf{B594} (2004) 283.


\bibitem{MRW}
  A.~D.~Martin, M.~G.~Ryskin and G.~Watt,
  Eur.\ Phys.\ J.\ C {\bf 44} (2005) 69

\bibitem{Brod}
  S.~J.~Brodsky, R.~Enberg, P.~Hoyer and G.~Ingelman,
  Phys.\ Rev.\ D {\bf 71}  (2005) 074020



\bibitem{dipole}  N.~N.~Nikolaev and B.~G.~Zakharov, Z. Phys. {\bf  C49} 
(1991) 607; Z. Phys. {\bf C53}  (1992) 331.


\bibitem{dipole2}  A.~H.~Mueller, Nucl. Phys.
{\bf B415}  (1994) 373; A.~H.~Mueller and B.~Patel, Nucl. Phys. {\bf B425}
(1994)  471.  


\bibitem{KL}
  Y.~V.~Kovchegov and E.~Levin,
  Nucl.\ Phys.\ B {\bf 577} (2000) 221

  
 \bibitem{KW}
  A.~Kovner and U.~A.~Wiedemann,
  Phys.\ Rev.\ D {\bf 64} (2001) 114002

\bibitem{W}
  M.~Hentschinski, H.~Weigert and A.~Schafer,
  Phys.\ Rev.\ D {\bf 73} (2006) 051501.


\bibitem{slopedif1}
  M.~B.~Gay Ducati, V.~P.~Goncalves and M.~V.~T.~Machado,
  Phys.\ Lett.\ B {\bf 506} (2001) 52
 
  
\bibitem{slopedif2}
  M.~B.~Gay Ducati, V.~P.~Goncalves and M.~V.~T.~Machado,
  Nucl.\ Phys.\ A {\bf 697} (2002) 767



\bibitem{MS}
  S.~Munier and A.~Shoshi,
  Phys.\ Rev.\ D {\bf 69} (2004) 074022



\bibitem{GM}
  K.~Golec-Biernat and C.~Marquet,
  Phys.\ Rev.\ D {\bf 71} (2005) 114005
      

\bibitem{charmdif}
  V.~P.~Goncalves and M.~V.~T.~Machado,
  Phys.\ Lett.\ B {\bf 588} (2004) 180

\bibitem{golec}
  K.~Golec-Biernat and A.~Luszczak,
  Phys.\ Rev.\  D {\bf 79} (2009) 114010


\bibitem{wusthoff}
  M.~Wusthoff,
  Phys.\ Rev.\ D {\bf 56} (1997)  4311

 \bibitem{nikqqg}
  N.~N.~Nikolaev and B.~G.~Zakharov,
 J.\ Exp.\ Theor.\ Phys.\  {\bf 78} (1994) 598
  [Zh.\ Eksp.\ Teor.\ Fiz.\  {\bf 105} (1994) 1117]; Z. Phys. {\bf  C64} 
(1994) 631;  N.~N.~Nikolaev, W.~Schaefer, B.~G.~Zakharov and V.~R.~Zoller,
  JETP Lett.\  {\bf 80} (2004) 371
  [Pisma Zh.\ Eksp.\ Teor.\ Fiz.\  {\bf 80} (2004) 423]

\bibitem{aktas}
  A.~Aktas {\it et al.}  [H1 Collaboration],
  Eur.\ Phys.\ J.\  C {\bf 48} (2006) 749

\bibitem{zeusdata}
S. Chekanov {\it et al.}  [ZEUS Collaboration],  Nucl.\ Phys.\  B {\bf 713} (2005) 3;  Nucl.\ Phys.\  B {\bf 800} (2008) 1.



\bibitem{difusivo}
  E.~Iancu, C.~Marquet and G.~Soyez,
  Nucl.\ Phys.\  A {\bf 780}  (2006) 52.

\bibitem{arata}
  A.~Hayashigaki,
  Nucl.\ Phys.\  A {\bf 775} (2006) 51.



\bibitem{dglap} V.N. Gribov and L.N. Lipatov, Sov. J. Nucl. Phys. {\bf 15} (1972) 438.
\bibitem{dglap1} G. Altarelli and G. Parisi, Nucl. Phys.  {\bf B126}  (1977) 298.
\bibitem{dglap2} Yu.L. Dokshitzer, Sov. Phys. JETP {\bf 46}  (1977) 641.




\bibitem{star}
J. Adams {\it et al.}  [STAR Collaboration],
  Phys.\ Rev.\ Lett.\  {\bf 97}  (2006) 152302.



\bibitem{cteq}
H.~L.~Lai {\it et al.}  [CTEQ Collaboration],
  Eur.\ Phys.\ J.\ C {\bf 12}, 375 (2000).

\bibitem{kkp}
  B.~A.~Kniehl, G.~Kramer and B.~Potter,
  Nucl.\ Phys.\ B {\bf 582}, 514 (2000).


















  




  
  


  


  

    
  








		 
		 





  
  

  
  





				



  
  

		  















		
			   



			   

			   





























\end{thebibliography}
\end{document}